\documentclass[runningheads]{llncs}
\usepackage{xspace}
\usepackage{graphicx}
\usepackage{float}
\usepackage{subfigure}
\usepackage{color}
\usepackage{booktabs}
\usepackage{marvosym}
\usepackage{amsmath}
\usepackage{enumitem}
\usepackage{listings}
\begin{document}
\title{An Empirical Analysis of Just-in-Time Compilation in Modern Databases
\thanks{A peer-reviewed version \cite{ADC2023_JIT} of this paper has been published at the Australasian Database Conference (ADC) 2023.}
}
%
%
\newcounter{myexample}[section]
\renewcommand{\themyexample}{\nthesection.\arabic{myexample}}
\newenvironment{myexample}{
     \refstepcounter{myexample}
     {\vspace{1ex} \noindent\bf  Example  \themyexample:}}{
     \vspace{1ex}} 


\newcounter{mydefinition}[section]
\renewcommand{\themydefinition}{\nthesection.\arabic{mydefinition}}
\newenvironment{mydefinition}{
     \refstepcounter{mydefinition}
     {\vspace{1ex} \noindent\bf  Definition  \themydefinition:}}{
     } 

\newcounter{mytheorem}[section]
\renewcommand{\themytheorem}{\nthesection.\arabic{mytheorem}}
\newenvironment{mytheorem}{\begin{em}
        \refstepcounter{mytheorem}
        {\vspace{1ex} \noindent\bf  Theorem  \themytheorem:}}{
        \end{em}} 

\newcounter{mylemma}[section]
\renewcommand{\themylemma}{\nthesection.\arabic{mylemma}}
\newenvironment{mylemma}{\begin{em}
        \refstepcounter{mylemma}
        {\vspace{1ex}\noindent\bf Lemma \themylemma:}}{
        \end{em}} 

		

\newcounter{operation}[section]
\renewcommand{\theoperation}{\nthesection.\arabic{operation}}
\newenvironment{operation}{\begin{em}
        \refstepcounter{operation}
        {\vspace{1ex}\noindent\bf Operation \theoperation:}}{
        \end{em}\eot\vspace{1ex}} 

\newcounter{observation}[section]
\renewcommand{\theobservation}{\nthesection.\arabic{observation}}
\newenvironment{observation}{\begin{em}
        \refstepcounter{observation}
        {\vspace{1ex}\noindent\bf Observation \theobservation:}}{
        \end{em}\eot\vspace{1ex}} 

\newcommand\blfootnote[1]{%
    \begingroup
    \renewcommand\thefootnote{}
\footnote{#1}%
    \addtocounter{footnote}{-1}%
    \endgroup
}

\newcommand{\proofsketch}{\noindent{\sc Proof Sketch: }}
\newcommand{\myproof}{\noindent{\bf Proof: }}

\newcommand{\nthesection}{\arabic{section}}
\newcommand{\eot}{\hspace*{\fill}\mbox{$\Box$}}
\newcommand{\eop}{\hspace*{\fill}\mbox{$\Box$}}




\newcommand{\stitle}[1]{\vspace{1ex} \noindent{{\bf #1}}}
\newcommand{\itemsection}[1]{\vspace{0ex} \noindent{\textbf{$\bullet$ ~ #1}}}
\newcommand{\sstitle}[1]{\vspace{1ex} \noindent{{\textit{ #1}}}}
\newcommand{\ssstitle}[1]{\vspace{1ex} \noindent{\textit{ #1}}}

\newcommand{\red}[1]{\color{red}{#1}}

\newcommand{\kw}[1]{{\textsf{#1}}\xspace}
\newcommand{\bkw}[1]{{\ensuremath {\mathsf{\textbf{#1}}}}\xspace}

\newcommand{\kwnospace}[1]{{\ensuremath {\mathsf{#1}}}}
\newcommand{\ltt}{\kw{LTT}}
\newcommand{\arr}{\kw{arrive}}
\newcommand{\vp}{\kw{p}}

\newcommand{\bellf}{{\sc Bellman-Ford}\xspace}
\newcommand{\bfalgo}{{\sc OR}\xspace}

\newcommand{\aalgo}{{\sc KDXZ}\xspace}

\newcommand{\dijk}{{\sc Dijkstra}\xspace}
\newcommand{\dalgo}{{\sc Two-Step-LTT}\xspace}

\newcommand{\dtalgo}{{\sc DOT}\xspace}

\newcommand{\genfunc}{{\sl timeRefinement}\xspace}
\newcommand{\pathc}{{\sl pathSelection}\xspace}
\newcommand{\fifo}{{\sl FIFO}\xspace}

\newcommand{\st}{starting time\xspace}
\newcommand{\sti}{starting-time interval\xspace}
\newcommand{\stsi}{starting-time subinterval\xspace}
\newcommand{\stsis}{starting-time subintervals\xspace}
\newcommand{\stis}{starting-time intervals\xspace}
\newcommand{\ti}{time interval\xspace}
\newcommand{\tis}{time intervals\xspace}
\newcommand{\ttime}{travel time\xspace}
\newcommand{\ttimea}{travel time\xspace}
\newcommand{\at}{arrival time\xspace}
\newcommand{\ata}{arrival-time\xspace}
\newcommand{\atf}{arrival-time function\xspace}
\newcommand{\ati}{arrival-time interval\xspace}
\newcommand{\ats}{arrival times\xspace}
\newcommand{\ed}{edge delay\xspace}
\newcommand{\eda}{edge-delay\xspace}
\newcommand{\edf}{edge-delay function\xspace}
\newcommand{\eds}{edge delays\xspace}
\newcommand{\wt}{waiting time\xspace}

\newcommand{\g}{\overline{g}}
\newcommand{\iend}{\tau}

\newcommand{\argmin}{\operatornamewithlimits{argmin}}

\newcommand{\myhead}[1]{\vspace{.05in} \noindent {\bf #1.}~~}
\newcommand{\cond}[1]{(\emph{#1})~}
\newcommand{\op}[1]{(\emph{#1})~}

\newcommand{\qc}{\ensuremath{Q^c}}

\newcommand{\rewrite}{\kw{XPathToReg}}

\newcommand{\upparen}[1]{\ensuremath{\mathrm{(}}{#1}\ensuremath{\mathrm{)}}}
\newcommand{\func}[2]{\funcname{#1}\upparen{\ensuremath{#2}}}
\newcommand{\funcname}[1]{\ensuremath{\mathit{#1}}}

\newcommand\AS{\textbf{as}\ }
\newcommand{\xsltsize}{\small}

\newcommand{\X}{{\cal X}}
\newcommand{\sem}[1]{[\![#1]\!]}
\newcommand{\NN}[2]{#1\sem{#2}}
\newcommand{\pcdata}{{\tt str}\xspace}

\newcommand{\exa}[2]{{\tt\begin{tabbing}\hspace{#1}\=\hspace{#1}\=\+\kill #2\end{tabbing}}}
\newcommand{\ra}{\rightarrow}
\newcommand{\la}{\leftarrow}
\newcommand{\rsa}{\_} 
\newcommand{\Ed}[2]{E_{{\scriptsize \mbox{#1} \rsa \mbox{#2}}}}
\newenvironment{bi}{\begin{itemize}
        \setlength{\topsep}{0.5ex}\setlength{\itemsep}{0ex}\vspace{-0.6ex}}
        {\end{itemize}\vspace{-1ex}}
\newenvironment{be}{\begin{enumerate}
        \setlength{\topsep}{0.5ex}\setlength{\itemsep}{0ex}\vspace{-0.6ex}}
        {\end{itemize}\vspace{-1ex}}
\newcommand{\ei}{\end{itemize}}
\newcommand{\ee}{\end{enumerate}}

\newcommand{\mat}[2]{{\begin{tabbing}\hspace{#1}\=\+\kill #2\end{tabbing}}}
\newcommand{\m}{\hspace{0.05in}}
\newcommand{\ls}{\hspace{0.1in}}
\newcommand{\beqn}{\begin{eqnarray*}}
\newcommand{\eeqn}{\end{eqnarray*}}

\newcounter{ccc}
\newcommand{\bcc}{\setcounter{ccc}{1}\theccc.}
\newcommand{\icc}{\addtocounter{ccc}{1}\theccc.}

\newcommand{\oneurl}[1]{\texttt{#1}}
\newcommand{\tabstrut}{\rule{0pt}{4pt}\vspace{-0.1in}}
\newcommand{\tabstruct}{\rule{0pt}{8pt}\\[-2ex]}
\newcommand{\stab}{\rule{0pt}{8pt}\\[-2.2ex]}
\newcommand{\sstab}{\rule{0pt}{8pt}\\[-2.2ex]}

\newcommand{\eat}[1]{}



\newcommand{\rdms}{{\sc rdbms}\xspace}
\newcommand{\sql}{{\sc sql}\xspace}
\newcommand{\dbms}{{\sc dbms}\xspace}

\newcommand{\cfig}{Figure~}
\newcommand{\ctab}{Table~}
\newcommand{\csec}{Section~}
\newcommand{\cdef}{Definition~}
\newcommand{\cthm}{Theorem~}
\newcommand{\clem}{Lemma~}
\newcommand{\cequ}[1]{Equation~(#1)}
\newcommand{\SG}{\mathbf{SG}}
\newcommand{\SA}{\mathbf{SA}}
\renewcommand{\AA}{\mathbf{AA}}

\newcommand{\xml}{{\sl XML}\xspace}
\newcommand{\xlink}{{\sl XLink}\xspace}
\newcommand{\xpath}{{\sl XPath}\xspace}
\newcommand{\xpointer}{{\sl XPointer}\xspace}
\newcommand{\rdf}{{\sl RDF}\xspace}
\newcommand{\tc}{{\sl TC}\xspace}
\newcommand{\dfs}{{\sl DFS}\xspace}
\newcommand{\DAG}{{\sl DAG}\xspace}
\newcommand{\DAGs}{{\sl DAG}s\xspace}
\newcommand{\grail}{{\sl GRAIL}\xspace}
\newcommand{\yesgrail}{{\sl Yes-GRAIL}\xspace}
\newcommand{\code}{\kw{code}}
\newcommand{\sit}{\kw{sit}}
\newcommand{\psit}{{\cal P}_{sit}}
\newcommand{\yescode}{{\sl Yes-Label}\xspace}
\newcommand{\nocode}{{\sl No-Label}\xspace}
\newcommand{\entry}{\kw{entry}\xspace}
\newcommand{\yngindex}{{\sl YNG-Index}\xspace}
\newcommand{\rqrun}{{\sl RQ-Run}\xspace}
\newcommand{\citeseerx}{{\sl citeseerx}\xspace}
\newcommand{\gouniprot}{{\sl go-uniprot}\xspace}
\newcommand{\uniprot}{{\sl uniprot150}\xspace}

\long\def\comment#1{}

\newcommand{\scc}{strongly connected component\xspace}
\newcommand{\sccs}{strongly connected components\xspace}
\newcommand{\sscc}{\kw{SCC}}
\newcommand{\ssccs}{\kwnospace{SCC}s\xspace}
\newcommand{\sccg}{\kwnospace{SCC}\textrm{-}\kw{Graph}}
\newcommand{\strongc}{\leftrightarrow}
\newcommand{\nstrongc}{\nleftrightarrow}
\newcommand{\emscc}{\kwnospace{EM}\textrm{-}\kw{SCC}}
\newcommand{\dfsscc}{\kwnospace{DFS}\textrm{-}\kw{SCC}}
\newcommand{\dfstree}{\kwnospace{DFS}\textrm{-}\kw{Tree}}
\newcommand{\len}{\kw{len}}
\newcommand{\dep}{\kw{depth}}
\newcommand{\tdep}{\kw{drank}}
\newcommand{\tlink}{\kw{dlink}}
\newcommand{\vedges}{up-edges\xspace}
\newcommand{\vedge}{up-edge\xspace}
\newcommand{\cvedge}{Up-Edge\xspace}

\newcommand{\drsscc}{\kwnospace{1P}\textrm{-}\kw{SCC}}
\newcommand{\drssccb}{\kwnospace{1PB}\textrm{-}\kw{SCC}}

\newcommand{\Bdrsscc}{\kwnospace{B}\textrm{-}\kwnospace{BR'}\textrm{-}\kw{SCC}}

\newcommand{\deprtree}{depth-ranked tree\xspace}
\newcommand{\cdeprtree}{Depth-Ranked Tree\xspace}
\newcommand{\drtree}{\kwnospace{BR}\textrm{-}\kw{Tree}}
\newcommand{\drplustree}{\kwnospace{BR}$^+$\textrm{-}\kw{Tree}}
\newcommand{\drscc}{\kwnospace{2P}\textrm{-}\kw{SCC}}
\newcommand{\updatedrank}{\kwnospace{update}\textrm{-}\kw{drank}}

\newcommand{\drtreeconstruct}{\kwnospace{Tree}\textrm{-}\kw{Construction}}
\newcommand{\drtreesearch}{\kwnospace{Tree}\textrm{-}\kw{Search}}
\newcommand{\depthrerank}{\kw{pushdown}}
\newcommand{\itrerank}{\kwnospace{iterative}\textrm{-}\kw{rerank}}
\newcommand{\drr}{\Downarrow}
\newcommand{\reach}{\kw{Rset}}
\newcommand{\earlyrejection}{\kwnospace{early}\textrm{-}\kw{rejection}}
\newcommand{\earlyacceptance}{\kwnospace{early}\textrm{-}\kw{acceptance}}
\newcommand{\greduce}{\earlyacceptance}
\newcommand{\drea}{\kwnospace{1P}\textrm{/}\kw{ER}}
\newcommand{\myinf}{\kw{INF}}

\newcommand{\gcloud}{\kw{GCloud}\xspace}
\newcommand{\degree}{\kw{Degree}\xspace}
\newcommand{\subgraph}{\kw{Subgraph}\xspace}
\newcommand{\pagerank}{\kw{PageRank}\xspace}
\newcommand{\bfs}{\kw{BFS}\xspace}
\newcommand{\keysearch}{\kw{KWS}\xspace}
\newcommand{\cc}{\kw{CC}\xspace}
\newcommand{\msf}{\kw{MSF}\xspace}
\newcommand{\dmax}{\kw{rmax}}
\newcommand{\mystar}{\kw{star}\xspace}
\newcommand{\twitter}{{\sl{Twitter-2010}}\xspace}
\newcommand{\friendster}{{\sl{Friendster}}\xspace}
\definecolor{lgray}{gray}{0.85}
\definecolor{llgray}{gray}{0.9}
\newcommand{\mycc}{CC\xspace}
\newcommand{\myccs}{CCs\xspace}
\newcommand{\mymsf}{MSF\xspace}
\newcommand{\oneroundmsf}{\kw{OneRoundMSF}}
\newcommand{\multiroundmsf}{\kw{MultiRoundMSF}}

\newcommand{\hashtomin}{\kw{HashToMin}\xspace}
\newcommand{\hashgreatertomin}{\kw{HashGToMin}\xspace}
\newcommand{\pramsimulation}{\kwnospace{PRAM}\textrm{-}\kw{Simulation}\xspace}

\newcommand{\pagerankpig}{\kwnospace{PageRank}\textrm{-}\kw{Pig}\xspace}
\newcommand{\bfspig}{\kwnospace{BFS}\textrm{-}\kw{Pig}\xspace}
\newcommand{\keysearchpig}{\kwnospace{KWS}\textrm{-}\kw{Pig}\xspace}

\newcommand{\ttwig}{\kwnospace{TwinTwig}\xspace}
\newcommand{\ttwigs}{\kwnospace{TwinTwig}s\xspace}
\newcommand{\ttjoin}{\kwnospace{TwinTwig}\kw{Join}}
\newcommand{\sdec}{\kwnospace{SDEC}\xspace}
\newcommand{\subgenum}{\kwnospace{SubgraphEnum}\xspace}
\newcommand{\mymap}{\kwnospace{map}\xspace}
\newcommand{\myreduce}{\kwnospace{reduce}\xspace}
\newcommand{\cascadejoin}{\kwnospace{Edge}\kw{Join}}
\newcommand{\starjoin}{\kwnospace{Star}\kw{Join}}
\newcommand{\multiwayjoin}{\kwnospace{Multiway}\kw{Join}}
\newcommand{\cost}{\kwnospace{cost}\xspace}
\newcommand{\mysize}{\kwnospace{card}\xspace}
\newcommand{\er}{\kwnospace{ER}\xspace}
\newcommand{\optdec}{\kwnospace{Optimal}\textrm{-}\kwnospace{Decomp}\xspace}

\newcommand{\ttone}{\kwnospace{TT1}\xspace}
\newcommand{\tttwo}{\kwnospace{TT2}\xspace}
\newcommand{\ttthree}{\kwnospace{TT3}\xspace}

\newcommand{\alEdge}{\kwnospace{Edge}\xspace}
\newcommand{\alMul}{\kwnospace{Mul}\xspace}
\newcommand{\alStar}{\kwnospace{Star}\xspace}

\newcommand{\alTTBO}{\kwnospace{TTBS}\xspace}
\newcommand{\alTTNLB}{\kwnospace{TTOA}\xspace}
\newcommand{\alTTLB}{\kwnospace{TTLB}\xspace}
\newcommand{\alTTFil}{\kwnospace{TT}\xspace}

\newcommand{\reffig}[1]{Figure~\ref{fig:#1}}
\newcommand{\refsec}[1]{Section~\ref{sec:#1}}
\newcommand{\reftable}[1]{Table~\ref{tab:#1}}
\newcommand{\refalg}[1]{Algorithm~\ref{alg:#1}}
\newcommand{\refdef}[1]{Definition~\ref{def:#1}}
\newcommand{\refthm}[1]{Theorem~\ref{thm:#1}}
\newcommand{\reflem}[1]{Lemma~\ref{lem:#1}}
\newcommand{\refcor}[1]{Corollary~\ref{cor:#1}}
\newcommand{\refex}[1]{Example~\ref{ex:#1}}
\newcommand{\refassum}[1]{Assumption~\ref{assum:#1}}
\newcommand{\refproperty}[1]{Property~\ref{property:#1}}
\newcommand{\refobv}[1]{Observation~\ref{obv:#1}}
\newcommand{\refproof}[1]{Proof~\ref{proof:#1}}

\makeatletter
\newcommand{\rmnum}[1]{\romannumeral #1}
\newcommand{\Rmnum}[1]{\expandafter\@slowromancap\romannumeral #1@}
\makeatother


\newcommand{\rebuild}{\kw{Rebuild}}
\newcommand{\BatchUpdatingAlgorithm}{\text{Batch Updating Algorithm}}
\newcommand{\BatchUpdatingProcessing}{\text{Batch Updating Processing}}
\newcommand{\BU}{\text{BU}}
\newcommand{\CO}{\text{Contraction Operation}}
\newcommand{\dynch}{\kw{DynCH}}
\newcommand{\dynchvcs}{\kw{DCH_{vcs}}}
\newcommand{\dynchvcb}{\kw{DCH_{vcb}}}
\newcommand{\dynchscs}{\kw{DCH_{scs}}}
\newcommand{\dynchscswinc}{\kwnospace{DCH_{scs}}\textrm{-}\kw{WInc}}
\newcommand{\dynchscswincp}{\kwnospace{DCH_{scs}}\textrm{-}\kw{WInc}}
\newcommand{\dynchscswincpp}{\kwnospace{DCH_{scs}^+}\textrm{-}\kw{WInc}}
\newcommand{\dynchscswincn}{\kwnospace{DCH_{scs}}\textrm{-}\kw{WIncDirect}}
\newcommand{\dynchscswdec}{\kwnospace{DCH_{scs}}\textrm{-}\kw{WDec}}
\newcommand{\dynchscswdecpp}{\kwnospace{DCH_{scs}^+}\textrm{-}\kw{WDec}}
\newcommand{\dynchscbwdec}{\kwnospace{DCH_{scb}}\textrm{-}\kw{WDec}}
\newcommand{\dynchscbwdecp}{\kwnospace{DCH_{scb}^+}\textrm{-}\kw{WDec}}
\newcommand{\dynchscbwdecn}{\kwnospace{DCH_{scb}}\textrm{-}\kw{WDecDirect}}
\newcommand{\dynchscbwinc}{\kwnospace{DCH_{scb}}\textrm{-}\kw{WInc}}
\newcommand{\dynchscbwincp}{\kwnospace{DCH_{scb}^+}\textrm{-}\kw{WInc}}
\newcommand{\dynchscb}{\kw{DCH_{scb}}}
\newcommand{\dynchscbn}{\kwnospace{DCH_{scb}}\textrm{-}\kw{Naive}}
\newcommand{\dynchscbp}{\kw{DCH^+_{scb}}}
\newcommand{\CHI}{\text{CH Index}\xspace}
\newcommand{\scgraph}{\text{$G_{sc}$}\xspace}
\newcommand{\RedOrd}{\text{reduce order}\xspace}
\newcommand{\ch}{\kw{CH}}
\newcommand{\gp}{{G'_{\oplus(e, k)}}}
\newcommand{\minw}{\kw{minWeight}}
\newcommand{\minwp}{\kw{minWeight^+}}
\newcommand{\ContractionHierarchies}{\text{Contraction Hierarchies}}
\newcommand{\RedPos}{\text{Pos}\xspace}
\newcommand{\ContractionHierarchy}{\text{Contraction Hierarchy}}
\newcommand{\shortcutindex}{\emph{shortcut index}\xspace}
\newcommand{\ssgraph}{\text{SS}\textrm{-}\text{Graph}\xspace}

\newcommand{\PartRelationshipEdge}{\text{part relationship edge}\xspace}
\newcommand{\UpdatingRelationshipGraph}{\text{updating relationship Graph}\xspace}
\newcommand{\OnLineAlgorithm}{\text{on-line query algorithm}\xspace}
\newcommand{\OffLineAlgorithm}{\text{index algorithm}\xspace}
\newcommand{\RoadUpdatingProcessing}{\text{road updating processing}\xspace}
\newcommand{\ShortcutRelationshipGraph}{\text{shortcut relationship graph}\xspace}
\newcommand{\SRGraph}{\text{SR-Graph}\xspace}
\newcommand{\cun}{\kw{CUN}}
\newcommand{\nbr}{\kw{nbr}\xspace}
\newcommand{\nbrin}{\kw{nbr^-}}
\newcommand{\nbrout}{\kw{nbr^+}}
\newcommand{\mydeg}{\kw{deg}}
\newcommand{\CompactShortcutRelationGraph}{\text{compact shortcut relation graph}\xspace}
\newcommand{\Effect}{\text{effect}\xspace}
\newcommand{\EffectSet}{\text{effect set}\xspace}
\newcommand{\GenerationSet}{\text{generation set}\xspace}
\newcommand{\DirectedRelation}{\text{directed relation}\xspace}
\newcommand{\UndirectedRelation}{\text{undirected relation}\xspace}
\newcommand{\RoadNetwork}{\text{road network}\xspace}
\newcommand{\DependentRelationship}{\text{dependent relationship}\xspace}
\newcommand{\AffectRelationship}{\text{affect relationship}\xspace}
\newcommand{\AffectRelationshipGraph}{\text{affect relationship graph}\xspace}
\newcommand{\CandidateRelationshipEdge}{\text{candidate relation edge}\xspace}

\newcommand{\ARG}{\text{$ARG$}\xspace}

\newcommand{\CalculateAffectSC}{\text{casc}\xspace}
\newcommand{\CalculateDependentSC}{\text{$cdsc$}\xspace}
\newcommand{\Calculation}{\text{Calculation}\xspace}
\newcommand{\DistancePreservedGraph}{\text{distance-preserved graph}\xspace}
\newcommand{\AffectCalculationSet}{\text{affected calculation set}\xspace}
\newcommand{\UnaffectCalculationSet}{\text{unaffected calculation set}\xspace}
\newcommand{\ACS}{\text{$\mathcal{ACS}$}\xspace}
\newcommand{\UCS}{\text{$\mathcal{UCS}$}\xspace}
\newcommand{\DependentCalculationSet}{\text{dependent calculation set}\xspace}
\newcommand{\DCS}{\text{$\mathcal{DCS}$}\xspace}
\newcommand{\CHNeighbourSet}{\text{CH neightbour set}\xspace}
\newcommand{\CHNS}{\text{$\mathcal{CHNS}$}\xspace}
\newcommand{\DirectedAffectCalculationSet}{\text{Directed Affect set}\xspace}
\newcommand{\DirectedCandidateSet}{\text{directed candidate set}\xspace}
\newcommand{\DACS}{\text{$\mathcal{DAS}$}\xspace}
\newcommand{\ShortcutCombination}{\text{shortcut combination}\xspace}
\newcommand{\CandidateSet}{\text{candidate set}\xspace}
\newcommand{\CS}{\text{$\mathcal{CS}$}\xspace}
\newcommand{\AffectShortcut}{\text{affect shortcut}\xspace}

\newcommand{\collectw}{\kw{collectWeight}}
\newcommand{\updatew}{\kw{updateWeight}}
\newcommand{\notifyw}{\kw{notifyWeight}}

\newcommand{\IndexWeightRecomputing}{\text{Index Weight Recomputing}\xspace}
\newcommand{\IWR}{\text{IWR}\xspace}
\newcommand{\QCHIU}{\text{Quick \CHI  Updating}\xspace}
\newcommand{\QU}{\text{QU}\xspace}
\newcommand{\QUP}{\text{QU$^+$}\xspace}
\newcommand{\RA}{\text{RA}\xspace}

\newcommand{\dsny}{\textit{NY}\xspace}
\newcommand{\dsco}{\textit{COL}\xspace}
\newcommand{\dsfl}{\textit{FLA}\xspace}
\newcommand{\dsca}{\textit{CAL}\xspace}
\newcommand{\dseus}{\textit{E-US}\xspace}
\newcommand{\dswus}{\textit{W-US}\xspace}
\newcommand{\dscus}{\textit{C-US}\xspace}
\newcommand{\dsus}{\textit{US}\xspace}

\newcommand{\topcaption}{%
 \setlength{\abovecaptionskip}{4pt}%
 \setlength{\belowcaptionskip}{0pt}%
 \caption}

\newcommand{\decb}{\kwnospace{DecB}\textrm{-}\kw{LMSD}}
\newcommand{\kecc}{\kw{kECC}}
\newcommand{\kec}{\kw{kEC}}
\newcommand{\keccs}{\kwnospace{kECC}s\xspace}
\newcommand{\ecc}{\kw{ECC}}
\newcommand{\eccs}{\kwnospace{ECC}s\xspace}
\newcommand{\inputgraph}{\textit{Input Graph G}\xspace}
\newcommand{\sparsifiedgraph}{\textit{Sparsified Graph G'}\xspace}
\newcommand{\preservedgraph}{\textit{kECC Preserved Graph G'}\xspace}
\newcommand{\observone}{\textit{Community Structure}\xspace}
\newcommand{\observtwo}{\textit{Core-Periphery Structure}\xspace}
\newcommand{\spanforest}{\kw{DisjointForest}}
\newcommand{\contract}{\kwnospace{CE}\textrm{-}\kw{Disk}}
\newcommand{\recover}{\kw{Recover}}
\newcommand{\componentid}{\kwnospace{CID}}
\newcommand{\kcore}{\kw{k}\textrm{-}\kw(core)}
\newcommand{\rand}{\kwnospace{Random}\textrm{-}\kw{Decom}}
\newcommand{\exact}{\kwnospace{Exact}\textrm{-}\kw{Decom}}
\newcommand{\bu}{\kwnospace{BU}\textrm{-}\kw{Decom}}
\newcommand{\td}{\kwnospace{TD}\textrm{-}\kw{Decom}}
\newcommand{\bottomup}{\kwnospace{Bottom}\textrm{-}\kw{Up}}
\newcommand{\hybrid}{\kwnospace{Hybrid}\textrm{-}\kw{Decom}}
\newcommand{\mydegree}{\kwnospace{degree}}
\newcommand{\topdown}{\kwnospace{Top}\textrm{-}\kw{Down}}
\newcommand{\hybrids}{\kw{Hybrid}}
\newcommand{\mem}{\kwnospace{Mem}\textrm{-}\kw{Decom}}
\newcommand{\id}{\kwnospace{id}}
\newcommand{\dsdblp}{\textit{DBLP}\xspace}
\newcommand{\dslj}{\textit{LiveJournal}\xspace}
\newcommand{\dsorkut}{\textit{Orkut}\xspace}
\newcommand{\dshw}{\textit{Hollywood}\xspace}
\newcommand{\dsuk}{\textit{uk-2005}\xspace}
\newcommand{\dsit}{\textit{it-2004}\xspace}
\newcommand{\dstwitter}{\textit{twitter-2010}\xspace}
\newcommand{\dssk}{\textit{sk-2005}\xspace}
\newcommand{\re}{\kwnospace{RE}\xspace}
\newcommand{\ce}{\kwnospace{CE}\xspace}
\newcommand{\rn}{\kwnospace{RN}}
\newcommand{\cn}{\kwnospace{CN}}
\newcommand{\kpg}{$k$\textrm{-}\kw{PG}}
\newcommand{\kpgs}{$k$\textrm{-}\kwnospace{PG}s\xspace}
\newcommand{\pg}{\kw{PG}}
\newcommand{\pgn}{\kwnospace{PG}}
\newcommand{\kpgn}{$k$\textrm{-}\kwnospace{PG}}
\newcommand{\remove}{\kwnospace{RE}\textrm{-}\kw{Disk}}
\newcommand{\contractmem}{\kwnospace{CE}\textrm{-}\kw{Mem}}

\newcommand{\VCS}{\text{DCH$_{vcs}$}\xspace}
\newcommand{\WDEC}{\text{DCH$_{scs}$-WDec}\xspace}
\newcommand{\WDECP}{\text{DCH$^+_{scs}$-WDec}\xspace}
\newcommand{\WINC}{\text{DCH$_{scs}$-WInc}\xspace}
\newcommand{\WINCP}{\text{DCH$^+_{scs}$-WInc}\xspace}
\newcommand{\SCB}{\text{DCH$_{scb}$}\xspace}
\newcommand{\SCBP}{\text{DCH$^+_{scb}$}\xspace}

\newcommand{\hyper}{\kw{HyPer}}
\newcommand{\mutable}{\kw{Mutable}}
\newcommand{\pgsql}{\kw{PostgreSQL}}

\author{
Miao Ma\inst{1}\and
Zhengyi Yang\inst{1}(\Letter) \orcidID{0000-0003-1772-6863}\and 
Kongzhang Hao\inst{1}\and
Liuyi Chen\inst{1}\and
Chunling Wang \inst{2}\and
Yi Jin\inst{2}
}
\authorrunning{M. Ma et al.}
%
\institute{The University of New South Wales, Sydney, Australia \\
\email{\{miao.ma,zhengyi.yang,kongzhang.hao,liuyi.chen\}@unsw.edu.au} \and
Data Principles (Beijing) Technology Co. Ltd, Beijing, China\\
\email{\{chunling.wang, yi.jin\}@enmotech.com}}
\maketitle              
\begin{abstract}
JIT (Just-in-Time) technology has garnered significant attention for improving the efficiency of database execution. It offers higher performance by eliminating interpretation overhead compared to traditional execution engines. LLVM serves as the primary JIT architecture, which was implemented in PostgreSQL since version 11. However, recent advancements in WASM-based databases, such as Mutable, present an alternative JIT approach. This approach minimizes the extensive engineering efforts associated with the execution engine and focuses on optimizing supported operators for lower latency and higher throughput. In this paper, we perform comprehensive experiments on these two representative open-source databases to gain deeper insights into the effectiveness of different JIT architectures.

\end{abstract}

\section{Introduction}


Databases are essential for storing, retrieving, and managing extensive data across industries like e-commerce, finance, healthcare, and logistics. The need for faster and more efficient data processing has led to research and innovation in enhancing database performance and optimizing resource use. An approach that has garnered significant interest is the incorporation of Just-in-Time (JIT) compilation into databases.

JIT compilation, a dynamic compilation approach, compiles code at run time just before execution, in contrast to Ahead-of-Time (AOT) compilation \cite{1960article}.
Early databases relied on AOT compilation result in limitations and significant overhead. JIT compilation, conversely, offers numerous advantages, including improved query execution speed, reduced memory usage, and adaptability to dynamic workloads. These advantages collectively enhance the query performance of JIT databases.
Due to these benefits, there has been a recent surge in the development of JIT databases focused on enhancing query performance. They can be primarily categorized into two types: LLVM (Low Level Virtual Machine)-based and WASM (WebAssembly)-based databases.
 
\sstitle{LLVM-based databases.} 
In LLVM-based databases, an intermediate representation (IR) is generated directly from the query execution plan, utilizing native LLVM or adapted LLVM. This IR is then compiled into machine code, undergoing effective optimization for faster compilation and improved query performance.
Notably, examples like Hyper\cite{neumann2011efficiently} and the open-source database \pgsql \cite{PostgreSQL} demonstrate LLVM-based JIT adoption. As Hyper is not an open source database, we've chosen \pgsql as the representative database system for LLVM JIT.
Starting from version $11$, \pgsql introduced JIT support, enabling on-the-fly compilation of \emph{expressions} within queries. These expressions are compiled into bytecode-based functions, replacing interpreted execution. This dynamic compilation generates optimized machine code tailored to the hardware architecture, enhancing query performance.


\sstitle{WASM-based databases.}
LLVM-based JIT databases outperform traditional databases but demand significant engineering efforts, including reengineering core compiler techniques. Given the evolving database landscape, continually reinventing JIT compilation in LLVM-based databases is costly and impractical.
To tackle this challenge, WASM-based databases provide an alternative. They aim to reduce the extensive and complex engineering work associated with the execution engine. A notable example is \mutable \cite{Mutable}, which introduces a novel query execution engine architecture. It employs WebAssembly as its intermediate representation (IR) and utilizes Google’s V8 and Binaryen as its backend for compilation and execution. This leverages Binaryen's fully parallel code generation and optimization capabilities. \mutable's innovative approach eliminates the need for extensive reengineering of techniques developed by the compiler community over decades.

%

\stitle{Aims and Contributions.} 
This paper conducts a thorough comparative analysis of two significant JIT databases, \pgsql and \mutable. We analyze their JIT implementations, evaluate their strengths and weaknesses, and identify potential areas for future JIT research in databases. Our contributions include:
%

\sstitle{(A) \underline{Comprehensive survey of JIT in modern databases.}} This study serves as a pioneering effort in surveying and comparing the performance of JIT compilation in contemporary open-source databases, with a particular emphasis on \pgsql and \mutable as prominent case studies. It stands as the first work to offer an in-depth analysis and comprehensive survey of well-established open-source JIT databases in modern database technology.


\sstitle{(B) \underline{Rigorous and extensive experimental studies.}} To investigate the performance of JIT in databases, we conducted in-depth experimental studies. Our experiments encompassed various aspects, including the evaluation of aggregation performance, a comparative analysis of grouping calculations, the examination of expression evaluation, the compiling time, and the assessment of performance with increasing data volumes. By designing comprehensive experiments and employing representative datasets, we ensured accurate and reliable results.

\sstitle{(C) \underline{A practical guidance for JIT compilation.}} This paper provides practical recommendations for JIT compilation grounded in empirical analysis and extensive comparisons of JIT database implementations. These guidelines constitute a valuable reference for both practitioners and researchers aiming to leverage the advantages of JIT compilation. Furthermore, we offer insights and stimulate future advancements in JIT databases, highlighting potential domains for further exploration and enhancement.

\stitle{Outline.} \refsec{pre} provides the background. \refsec{related} reviews the related works. \refsec{impl} presents the details about JIT implementation in \pgsql and \mutable. \refsec{exp} evaluates and analyzes the performance of \pgsql and \mutable. \refsec{future} discusses the future works and \refsec{conclusion} concludes the paper.

\section{Preliminary}
\label{sec:pre}

In this section, we provide an overview of key concepts, including ahead-of-time compilation, just-in-time compilation, and the TPC-H Benchmark.

\stitle{AOT Compilation.} AOT (Ahead-Of-Time) compilation is commonly associated with programming languages like C and C++. In AOT, programs can execute without a runtime and can directly link static or dynamic libraries to binary code. One advantage of AOT is to eliminate the need for the time and overhead associated with online compilation. This advantage is particularly beneficial for short-running programs with relatively flat method hotness curves \cite{AOTvsJIT}. AOT allows for code analysis and optimization without considering resource costs caused by ahead-of-time compilation. However, AOT typically lacks access to reliable profile data which is limiting its effectiveness. AOT also cannot perform certain optimizations available in JIT, such as run time profile-guided optimization, pseudo-constant propagation, or indirect-virtual function inlining.

\stitle{JIT Compilation.}
JIT compilation is a dynamic compilation technique that has gained significant attention in various domains. Unlike traditional AOT compilation, JIT compilation occurs at run time, generating optimized machine code just before it is executed. By fast compilation and dynamically optimizing code based on run time information, JIT compilation surpasses statically compiled or interpreted code in performance. It can the Intermediate language\cite{IL} code by translating to the native code repeatly when it is supposed to be executed\cite{METULA20113}. It has become an integral part of modern software systems especially in VM and web browsers. In the realm of database systems, JIT compilation holds promise for enhancing query execution performance. 

\stitle{TPC-H Benchmark.}
The TPC-H benchmark\cite{tpch} is widely recognized for evaluating database system performance. It simulates real-world analytical scenarios with complex operations like multi-table joins and aggregations. Adhering to TPC-H guidelines ensures fair comparisons across systems. The evaluation focuses on metrics like query execution time and throughput, providing insights into query performance and concurrent workload handling. We use TPC-H to assess \pgsql and \mutable, aiming to understand JIT compilation's impact on complex query performance.

\section{Related Work}
\label{sec:related}

\stitle{LLVM.}
With the development of modern CPUs, the most widely used database execution framework of Volcano iterators \cite{Volcano} exposes the limitations of Cache inefficiency, CPU inefficiency, iterator overhead. In 2005, Hyper-Pipelining Query Execution proposes a vectorized model which can improve CPU execution efficiency. However, it also exposed two dangers including the significant fraction thereof and mispredicted branch \cite{Hyper-Pipelining}. In 2010, the paper \cite{holistic} customized code generation to optimize it holistically by generating the hardware-specific source code. In 2011, \cite{neumann2011efficiently} introduced dynamic compilation and parallel execution framework, highlighting the advantages of utilizing LLVM for generating cross-platform Intermediate Representation (IR) code. This approach exhibited rapid compilation and robust code optimization while ensuring compatibility with existing C++ code through features like function calls and direct memory access.
Subsequently, numerous database systems which were inspired by HyPer \cite{Hyper} have adopted similar dynamic compilation techniques. In 2014, the concept of Morsel-driven parallelism \cite{leis2014morsel} was introduced to address performance bottlenecks stemming from underutilized system resources when utilizing a single core. This concept involved dividing query execution data into morsels and implementing an operator pipeline with dynamic scheduling strategies. A morsel pipeline was treated as an independent task and executed by a worker in a thread pool. This approach enabled fine-grained dynamic adjustments, maximized throughput, and improved cache locality.
In 2018, further optimization in compilation based on the Morsel-driven idea was explored \cite{kohn2018adaptive}. This strategy entailed collecting statistical information during query execution, estimating compilation time based on the collected statistics, and dynamically switching to compilation mode. By allowing different query groups to employ distinct execution modes, more efficient query execution was achieved.


\stitle{WASM.} Recently, \mutable \cite{haffner2023simplified} offers an alternative JIT approach that could minimize the complex engineering work required by the execution engine, while it could also optimize supported operators to achieve lower latency and higher throughput. It utilizes V8 and Binaryen as the backend for compilation and execution. It takes advantage of Binaryen's fully parallel code generation and optimization capabilities which will maximize CPU utilization.\pgsql has also embraced JIT based on LLVM, but there is no performance comparison analysis conducted thus far between these two representative JIT architecture of LLVM and WebAssebly, namely \pgsql and \mutable.

\stitle{Others.} 
Various JIT compilers are utilized in different programming languages and platforms to enhance code execution efficiency. In the Python ecosystem, JIT compilers like PyPy, Numba, and Cython aim to optimize Python code execution, with Numba even offering the option to disable the Global Interpreter Lock (GIL) \cite{python}. LuaJIT \cite{LUAJIT} is a trace-based JIT compiler designed for the Lua programming language, generating efficient code to boost Lua program execution. Meanwhile, in Java, Java Hotspot stands out as an efficient JIT compiler for the Java Virtual Machine (JVM). It identifies the frequently used methods and optimizes them into Java bytecode, with JDK-9 \cite{inproceedings} introducing two compilers, \textit{c1} and \textit{c2}, to cater to different optimization needs. While \textit{c1} prioritizes speed, \textit{c2} applies an array of optimizations
to produce high-quality code and optimize program execution \cite{AOTvsJIT}.

\section{JIT Implementations}
\label{sec:impl}
\begin{figure}[t]
\centerline{\includegraphics[width=.8\textwidth]{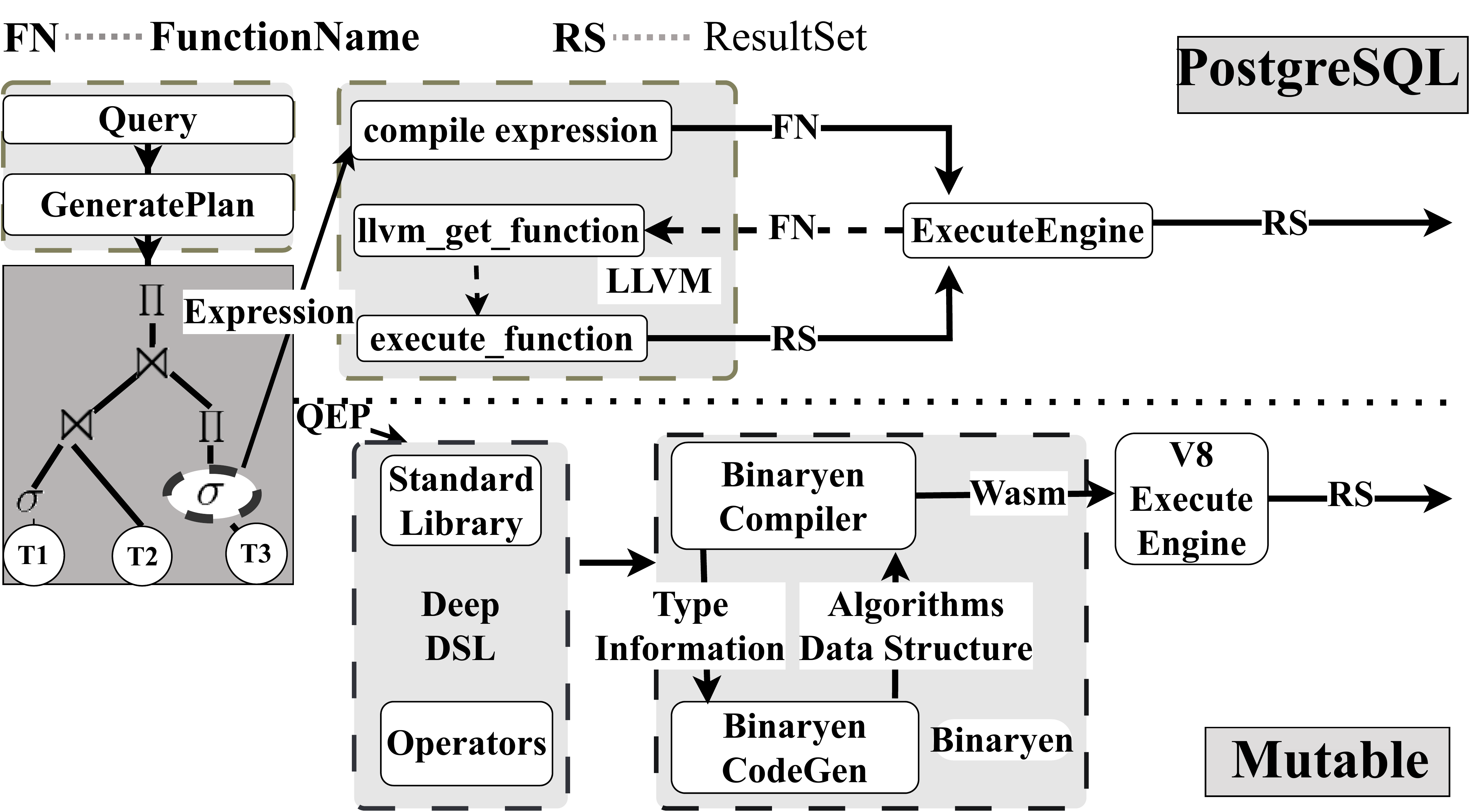}}
\caption{The JIT architecture of \mutable and \pgsql}
\label{fig1}
\end{figure}

In this section, we delve into the JIT implementation of both \pgsql and \mutable, and their architectures are illustrated in Fig. \ref{fig1}.

\subsection{\pgsql}

Starting from version 11, \pgsql introduces JIT compilation through an LLVM-based approach. The primary focus of JIT implementation centers on optimizing query \emph{expressions}. This is achieved by incorporating the corresponding JIT provider, registered as an external dependency library, into \pgsql's execution engine.
The default JIT provider in \pgsql is \textit{llvmjit} \cite{LLVMJIT}. However, \pgsql offers a JIT provider registration framework, enabling the replacement of JIT providers. This design accommodates the future development of provider libraries while ensuring the physical separation of provider code from the database execution engine.
Once JIT is enabled, \pgsql initiates JIT for expressions. The process involves compiling the query's expressions into bytecode-based functions. These generated functions are seamlessly integrated into the evaluation function of \textit{ExprState}, facilitating JIT incorporation into the existing execution framework without the need for modifications to the original SQL execution process.



The JIT implementation in \pgsql can be categorized into three main aspects: (1) JIT Compiled Expression, (2) Inlining, and (3) Optimization Passes. 

\stitle{JIT Compiled Expression.} Modern databases, particularly those focused on online analytical processing (OLAP), frequently encounter CPU performance bottlenecks during expression calculations and table tuple scanning. To tackle these challenges, \pgsql harnesses the inherent capabilities of LLVM for accelerated expression operations. This encompasses optimizing expression evaluations and tuple deformation processes, which eliminate unnecessary function calls. Additionally, tuple deformation optimization involves the conversion of disk-based tuples into an in-memory state, reducing I/O costs and enhancing cache utilization, thereby significantly improving overall performance.


\stitle{Inlining.} In \pgsql, there are numerous redundant copies of general utility functions. Rewriting these functions is both impractical and undesirable. To address this, \pgsql utilizes inlining techniques, such as loop flattening \cite{loopflatten}, which could consolidate the nested loops into a single loop. This consolidation eliminates redundant jumps and unnecessary code branching during execution. Inlining effectively removes the overhead associated with function boundaries, significantly reducing code size and enhancing the overall efficiency of the database engine. 


\stitle{Optimization by Passes.}  In \pgsql, LLVM's passes are employed to further optimize the generated code in pipelines. Two core passes in LLVM are utilized: Analysis pass and Transform pass. The Analysis pass calculates statistical information of IR units, aiding in debugging and display. Conversely, the Transform pass modifies and optimizes IR into a simpler and equivalent form. For example, by analyzing the relationship of function calls using strongly connected components, optimization passes can eliminate redundant paths and code. Furthermore, the vectorization transform pass maximizes the utilization of vector registers.
However, it's important to note that these passes come with associated costs. If their cost exceeds the execution cost of the query, they can potentially slow down the execution. To mitigate this issue, \pgsql employs a cost threshold to control the use of optimization passes.



\subsection{\mutable}

Mutable \cite{Mutable} is an experimental in-memory database designed to implement an alternative JIT architecture with the goals of improving throughput, reducing response time, and simplifying programming workloads.
To achieve these objectives, Mutable adopts the WasmV8 architecture \cite{Mutable}, which combines Binaryen and Google's V8 as the JIT backend.
In the architecture of Mutable (Fig. \ref{fig1}), the Query Execution Plan (QEP) is fully compiled by the Binaryen compiler. Subsequently, Binaryen \textit{CodeGen} generates WebAssembly code, which is then handed over to the embedded V8 engine for execution. 


\stitle{Fully Compiled QEP.} 
Mutable adopts a fully compiled JIT approach as its backend execution engine. This allows for the implementation of operators in a deep Domain Specific Language (DSL) separately, resulting in a fully compiled Query Execution Plan (QEP). This approach provides a global view for optimizing instructions within the system. Additionally, Mutable utilizes an iterator architecture implemented through an inheritance framework, offering flexibility and abstraction structure which is similar to the Volcano iterator \cite{graefe1994volcano}. Operators can be effortlessly added to the framework by overriding the \textit{(open()-next()-close())}execution functions, ensuring seamless integration into the execution framework. 


\stitle{Standard Library Implementations.} 
WebAssembly lacks of a standard library \cite{haffner2023simplified}. To overcome this limitation, Mutable generates standard functions, like hash functions and item comparison functions, using a deeply embedded domain-specific language (deep DSL) \cite{Mutable}. This approach allows for fine-grained definition. However, it's important to note that Mutable produces monomorphic code that is specific to  query execution in contrast to supporting the entire C standard library (LIBC) or Standard Template Library (STL).



\stitle{Binaryen in WasmV8.} 
WasmV8 utilizes Binaryen as the \textit{CodeGen} framework to generate WebAssembly code. It employs WASM-specific optimizations and produces compacted data structures through various passes of optimization\cite{Binaryen}. Binaryen also extracts morsel units for pre-compiled libraries in a pipeline. Additionally, it incorporates lazy compilation to decouple function dependencies in a global view, reducing the burden of building shaders. This approach allows for an early launch of the execution engine, which in turn reduces total execution time. Moreover, Binaryen achieves faster compilation by separating the tasks of work threads and allowing some worker threads to handle prediction instructions or independent tasks.


\stitle{Adoptive Compilation in V8.} 
Adaptive compilation in \mutable is facilitated by the V8 engine, which includes \textit{TurboFan} and \textit{Liftoff}. TurboFan utilizes a graph-based intermediate representation (IR) and implements various optimizations, including strength reduction, inlining, code motion, instruction combining, and sophisticated register allocation. It operates closely to machine code, bypassing several stages to achieve efficient performance \cite{V8}. In contrast, Liftoff serves as a baseline compiler specifically tailored for WebAssembly. Its primary objective is to minimize startup time for WASM-based applications by rapidly generating code \cite{V8}. In mutable, Liftoff is typically disabled by default, with an emphasis on alternative optimization strategies for JIT compilation.


\section{Experimental Evaluation}
\label{sec:exp}


Our aim is to evaluate and compare the performance of the JIT architectures in \pgsql and \mutable using the TPC-H benchmark. In order to isolate the effects of parallelism, we disabled parallel execution and enabled JIT in \pgsql by default. 
Due to the current limitations of WebAssembly's 32-bit addressing, we conducted our experiments using a TPC-H data scale of $1G$. As \mutable currently supports only a subset of TPC-H queries (Q1, Q3, Q6, Q12, and Q14), we based our initial experiment on the performance of aggregation using Q1, analyzed grouping performance using Q3, and evaluated expression calculation performance using Q6, and overall performance using Q12 and Q14. Additionally, we compared the performance of JIT with and without parallel execution in \pgsql.

\stitle{Environments.} We utilized \pgsql 15.2, which has undergone extensive optimization for OrderBy and GroupBy operations. The default JIT provider is \textit{llvm} \cite{LLVMJIT}, was employed for \pgsql JIT. For \mutable, we compiled a release version from version 0.17. We conduct our tests on Intel(R) Xeon(R) Gold 6342 CPU @ 2.80GHz server with two physical CPUs and 24 cores.

\stitle{Compared Strategies.} To ensure results accurate, each experiment was repeated five times, and the median value was obtained. To eliminate the disk I/O cost in \pgsql, we set the size of buffer pool to $10G$ and discard the first test result in \pgsql.


\renewcommand\thesubfigure{\arabic{subfigure}}
\begin{figure}[t]
\centering 
\subfigure[Execution time of \mutable and \pgsql in TPC-H 1G dataset]{
\label{Fig.sub.1}
\includegraphics[width=.475\textwidth]{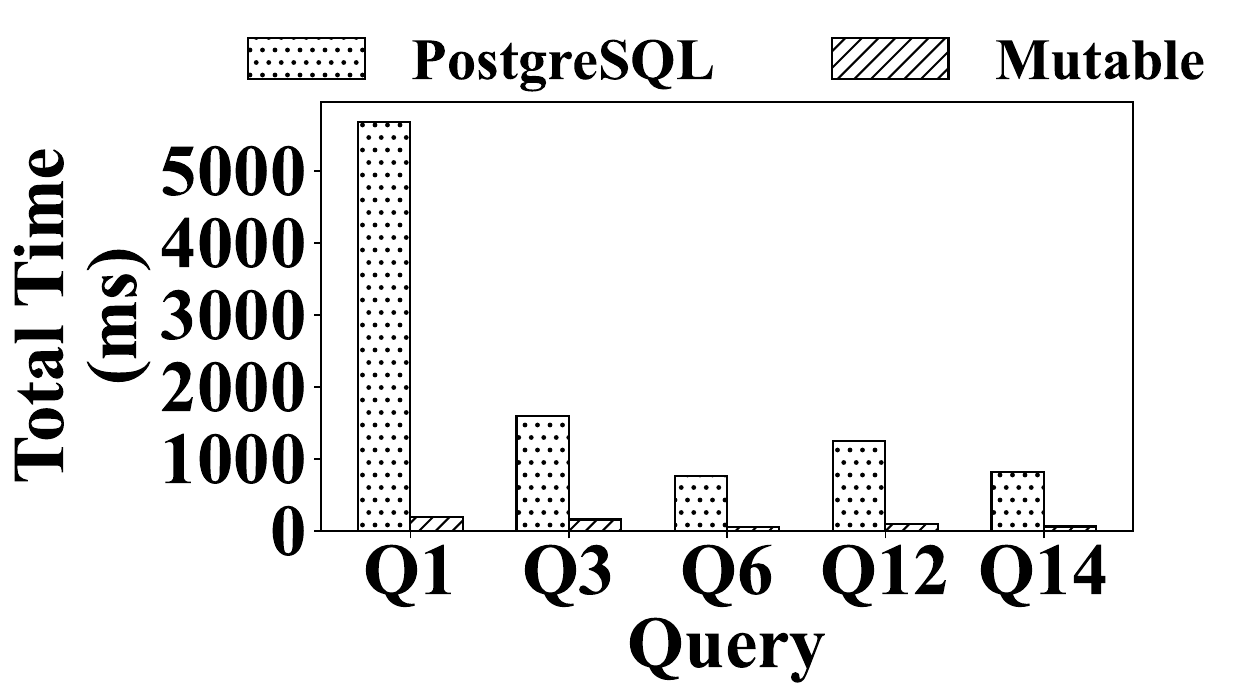}}
\subfigure[Compile time of \mutable and \pgsql in TPC-H 1G dataset]{
\label{Fig.sub.2}
\includegraphics[width=.46\textwidth]{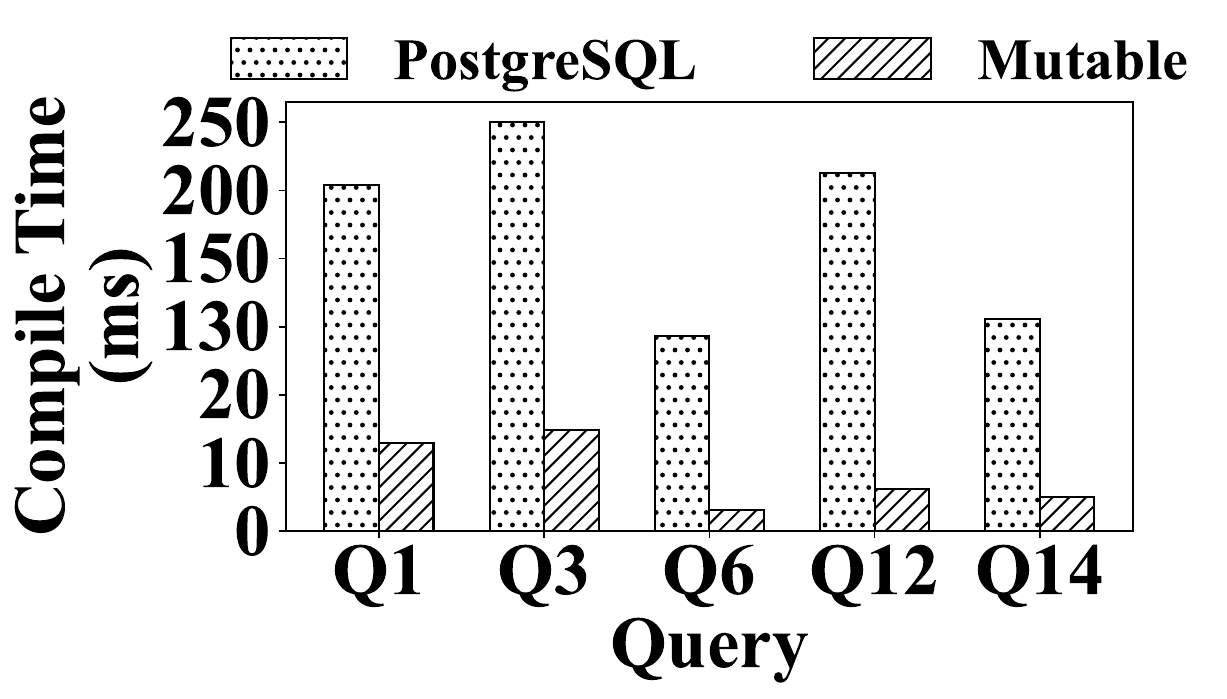}}
\\
\subfigure[Execution time of \pgsql in TPC-H 1G dataset]{
\label{Fig.sub.3}
\includegraphics[width=.445\textwidth]{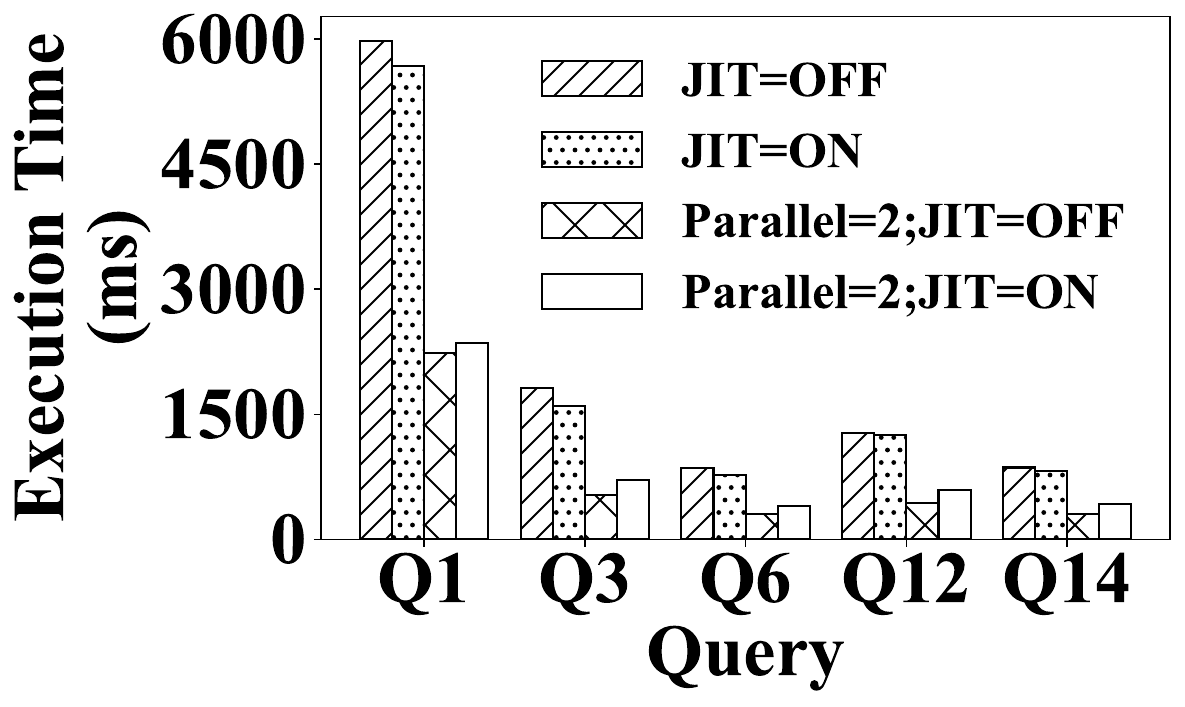}}
\subfigure[Speedup rate of JIT for parallel execution in \pgsql]{
\label{Fig.sub.4}
\includegraphics[width=.47\textwidth]{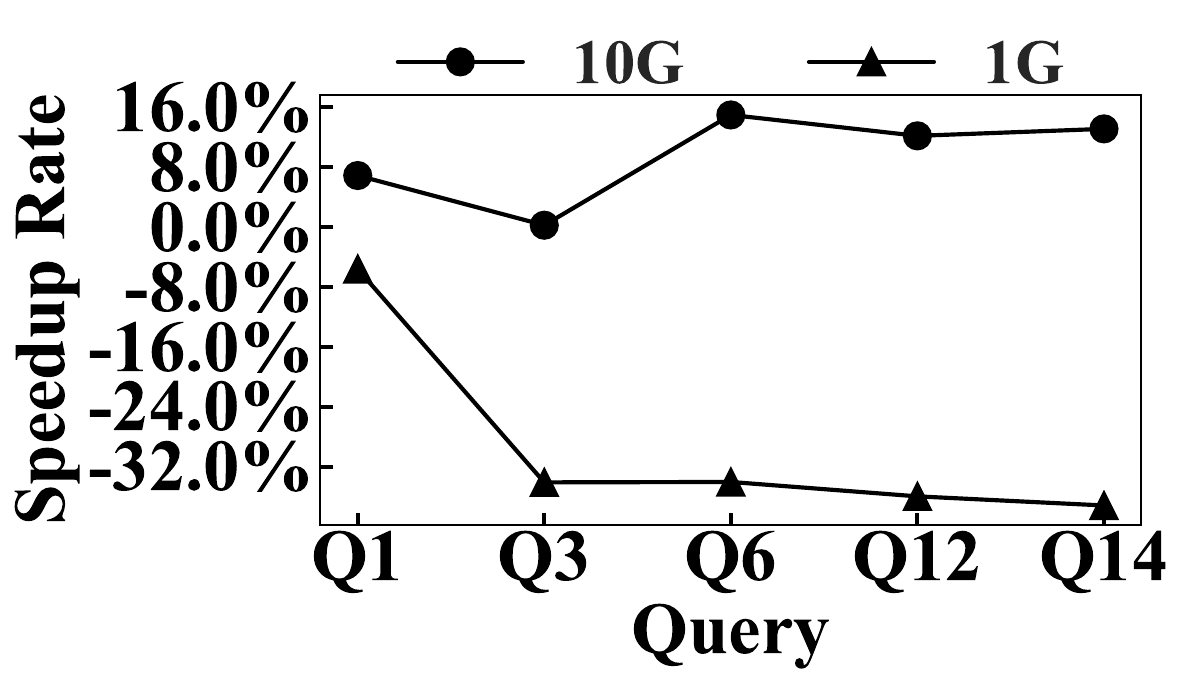}}
\caption{Experiment Results}
\label{1}
\end{figure}


\stitle{Exp-1: Overall Comparison.} 
Fig. \ref{1}.1 illustrates the performance comparison between \pgsql and \mutable based on the same query plan, where \mutable outperforms \pgsql in all queries, by an average of $16.11$ times.
The most significant difference is evident in Q1, where \mutable outperforms \pgsql by $27.75$ times. 
We analyse the performance for each query in the following.

\sstitle{Q1.} Q1 is distinguished by its CPU-intensive nature, involving eight aggregation functions \cite{DeepinTPCH}. Both \mutable and \pgsql employ an identical query plan for Q1, comprising \textsc{TableScan}, \textsc{HashGroup}, and \textsc{Sort} operations.
Despite both utilizing JIT compilation for generating aggregation functions, \mutable consistently outperforms \pgsql. This can be primarily attributed to three key optimizations within \mutable.
Firstly, \mutable integrates the aggregation functions using Webassembly's DSL (Domain-Specific Language), enabling the database-specific optimized code in compiled Query Execution Plans (QEPs).
Secondly, \mutable customizes the framework for aggregation code generation, reducing the generation of unnecessary code. Specifically, \mutable generates only the necessary functions for a given query, depending on the required aggregations.
Thirdly, \mutable supports the full pipeline model of compiling SQL queries into executable code. This involves grouping the pipeline breaker \cite{neumann2011efficiently}, allowing the decomposition of jobs into several sub-processes. Each sub-process can execute efficiently, benefiting from improved cache locality.

\sstitle{Q3.} Q3 involves a join of multiple tables with a large number of intermediate results, along with grouping, sorting, and aggregation operations. In Fig. \ref{Fig.sub.1} shows that \mutable outperforms \pgsql by a factor of $10.47$ in Q3.
%
%
The performance advantage of \mutable in this query is due to its JIT optimizations on Hash-based grouping. Typically, a \textsc{HashTable} needs to generate complete instructions for all data types, resulting in potentially redundant instructions. Additionally, when collisions occur during the hash value lookup, it necessitates a minimum of $n$ callbacks to resolve $n$ keys \cite{haffner2023simplified}. In contrast, \mutable generates specialized \textsc{HashTable} implementations for each query, with all hash table operations fully \emph{inlined} into the query code \cite{haffner2023simplified}. This approach integrates a proprietary \textsc{HASH} algorithm directly into the compiled Query Execution Plans (QEPs) to reduce function calls.
\mutable also fully inline comparison functions to minimize context switches, leading to improved query performance.

\sstitle{Q6.} 
Q6 is a single table query operation with aggregate operations and sequential scans. From Fig. \ref{Fig.sub.1}, \mutable outperforms \pgsql by $15.32$ times, primarily because of the following factors.
Firstly, Q6 extensively employs conjunctions in the \textit{WHERE} condition and if-condition blocks \cite{boncz2014tpc}. These conjunctions can be optimized through branch prediction in JIT compilation by converting control dependence into data dependence, therefore enhancing performance \cite{BranchPrediction}. 
In contrast to \pgsql, where expressions are delegated to LLVM in a serial manner, \mutable compiles all conjunction expressions together within a single filter operator. This allows for short-circuiting and constant folding optimizations to be applied, effectively trimming and optimizing expressions.
In addition, \mutable also extends the type of unreachable code to perform code elimination when code becomes unreachable using the control flow graph to further optimize the code.
This query demonstrates \mutable's performance improvement not only in complex queries but also in simpler ones.


\sstitle{Q12 \& Q14.}
Q12 and Q14 both involve Join operations. Q12 includes general grouping, sorting, and aggregation operations, while Q14 is a simpler query with only join and aggregations. In both cases, \mutable outperforms \pgsql significantly, with improvements of $13.32$ and $13.69$ times, respectively.
These queries underscore the advantages of \mutable in handling join, grouping, aggregation, and full Query Execution Plan (QEP) compilation, which jointly contribute to the overall performance improvement.

%



\stitle{Exp-2: Compile Time.} 
Fig. \ref{1}.2 highlights a significant difference in compile times between \pgsql and \mutable across all test cases. \mutable has an average compile time of only $8.4$ms, which outperforms \pgsql by $26$ times.
The shorter compilation time of \mutable is mainly attributed to Binaryen, which employs compact data structures during compilation and facilitates complete parallel compilation and optimization. 
Binaryen's efficiency is further enhanced by reusing existing memory blocks through its arena allocator for each WebAssembly module, reducing memory allocation costs and enhancing data cache utilization. Additionally, Binaryen implements an efficient in-place traversal of the abstract syntax tree (AST) during the compilation stage, considerably improving traversal speed.

\stitle{Exp-3: Parallel Execution vs JIT.} Fig. \ref{1}.3 indicates the query execution time for \pgsql with JIT and parallel execution.
It is evident that enabling JIT in \pgsql increases query performance by approximately $5\%$ when parallel execution is turned off. However, when the number of threads is set to two in parallel execution, JIT has a negative impact on query performance. This is because of the significant overhead of serial compilation in \pgsql's JIT design, especially for small datasets or simple queries. 
Note that \mutable, which does not support parallel execution, was not included in this experiment.


\stitle{Exp-4: Varying Datasets for Parallel Execution.} 
For parallel execution, which is enabled by default in \pgsql 10 and later versions, we conducted additional testing of JIT across datasets of varying sizes. Fig.\ref{1}.4 illustrates the speedup achieved by JIT in \pgsql in different datasets, specifically $1G$ and $10G$, each with two parallel workers.
The results indicate that JIT in \pgsql has a negative impact on performance when using a $1G$ dataset. However, as the dataset size increases while the compilation time remains relatively stable, the proportion of compilation time within the query execution time gradually decreases. Consequently, JIT in \pgsql exhibits a positive effect in the $10G$ TPC-H dataset test. With the larger dataset, the improvement brought by JIT execution becomes more pronounced, and the benefits surpass the compilation overhead.
These findings further corroborate the insights gained from Exp-3. Nevertheless, it is important to note that due to the inherent limitation of 32-bit addressing in WebAssembly, which restricts dataset sizes to under $4G$, we were unable to conduct tests on larger datasets using \mutable.

\section{Discussions and Future Work}
\label{sec:future}

\begin{table}[tb]
    \centering
    \caption{The comparison of \pgsql and \mutable in JIT}
\label{table1}
\small
\begin{tabular}{ |c|c|c| } 
 \hline
     & \pgsql & \mutable \\
  \hline
 Framework & LLVM & Webassembly V8 \\
  \hline
    JIT Unit & Expression & QEP \\
  \hline
 Adaptive Execution & No & Yes \\
  \hline
  Debug Friendly & No & Yes \\
    \hline
   Maturity & Yes & No \\
  \hline
   Parallel Compilation & No & Yes \\
  \hline
\end{tabular}
\label{table2}
\end{table}


\stitle{General Comparison.} 
After conducting and analyzing various tests between the two databases, we have summarized their respective features in terms of JIT compilation, as presented in Table \ref{table2}.
For the JIT framework, \pgsql opts for LLVM as the JIT provider, leveraging the advantages of mature LLVM technology and an active community.
In contrast, \mutable utilizes the existing WebAssembly and V8 frameworks, complemented by a custom DSL to implement its JIT module.

JIT compilation in \pgsql primarily emphasizes compiling \emph{expressions} rather than fully compiling QEPs which leads to significant limitations of JIT in \pgsql.
\mutable enables the complete compilation of QEPs, allowing for comprehensive optimization, effective compiled code, rapid parallel code generation, and resource-friendly caching of compiled code. As a result, \mutable consistently achieves significantly better near-native execution speed compared to \pgsql.

Moreover, it is worth noting that \pgsql lacks support for adaptive execution \cite{Hyper}, a feature that enables dynamic selection among different JIT modes. In contrast, \mutable utilizes V8 to implement adaptive execution, incorporating a tracing JIT compiler that can dynamically switch between JIT and interpreter modes. Meanwhile, \pgsql relies on a user-defined threshold to control optimization passes. This distinction highlights the flexibility and adaptability of \mutable's JIT framework.

In terms of debugging, JIT-compiled code can be highly optimized and challenging to correlate with the original source code, making visualization of the debugging process crucial. \pgsql often requires the use of external tools like LLDB \cite{LLDB} or GDB \cite{GDB} to disassemble the code for analysis. However, \mutable offers a more user-friendly debugging experience by integrating with Chrome's DevTools. This integration allows for better visualization of debugging information encoded in the WebAssembly file, supports profiling of WebAssembly code, and facilitates memory inspection. These features provide \mutable with a significant advantage in terms of visual debugging and code analysis.

For parallel compilation, while LLVM 8.0 in \pgsql currently lacks parallel compilation capabilities, the V8 engine used in \mutable is capable of parallel compilation, giving \mutable an advantage in terms of compilation efficiency. An interesting future work could involve updating the LLVM version in \pgsql to a higher version that supports parallel compilation.



\stitle{Future Work.}
We propose some future directions to further explore JIT technology and contribute to JIT in the database field.

\begin{itemize}[leftmargin=*]
\item \underline{\textit{Integrating WASM into \pgsql.}} We plan to improve \pgsql's JIT performance by implementing WASM-based JIT. This involves starting with expression-level WASM JIT without altering the engine's structure. Upon successful validation, we will extend JIT compilation to encompass full QEPs. 
Additionally, we aim to enable adaptive execution in \pgsql by integrating the adaptive execution tracing compiler from V8 to dynamically adjust execution plans based on statistical information.




\item \underline{\textit{Expanding Operators in \mutable.}} Our second focus is on improving \mutable's JIT performance and debugging visibility. 
This includes iteratively optimizing JIT operator capabilities, integrating complex operators like \textsc{Semi-Join}, \textsc{With Rollup}, and \textsc{Window} functions.
Priority will be given to enhancing CPU-intensive operators through strategic JIT function replacements. Our objectives also include improving observability and addressing concurrency architecture. Furthermore, we plan to gather runtime data during the JIT process to provide further optimization based on resource utilization, memory, threads, processing times, and more.

\item \underline{\textit{Harnessing Novel Hardware.}} We aim to leverage WebAssembly's extended capabilities to address the 32-bit addressing constraint in \mutable, enabling performance evaluation with larger data sizes. Additionally, we will utilize the WebAssembly SIMD (Single Instruction Multiple Data) Extension \cite{SIMD,FusedTableScans} to facilitate intra-core parallelization, optimizing performance by consolidating aggregation primitives into a cohesive loop.

\item \underline{\textit{Expanded Practical Evaluation.}} Our analysis will cover a broader spectrum, from single-core to multi-core scenarios, with datasets of varying size. We aim to include more JIT-enabled databases such as Hyper \cite{HyPerArchitecture}, Umbra \cite{Umbra}, MonetDB \cite{MonetDB}, and others. 
Our goal is to distill and present prevalent performance patterns seen in SQL queries within the TPC-H benchmark. 
We will assess various operator implementations, considering both multithreaded and SIMD contexts. This comprehensive approach aims to provide insights into performance dynamics across different products and configurations.

\end{itemize}

\section{Conclusion}
\label{sec:conclusion}
In summary, the research focuses on analyzing the performance of two open-source databases: \pgsql and \mutable. These databases employ different JIT compilation approaches, namely LLVM and WebAssembly. \pgsql integrates LLVM-based JIT with mature optimizations and compatibility, allowing users to enable or disable JIT functionality as needed. While \mutable inherit the advantages of WebAssembly. In experiments, \mutable outperformed \pgsql significantly. This research enhances our understanding of JIT in databases, providing valuable insights for practitioners and researchers seeking to optimize database systems for improved performance.


\subsubsection*{Acknowledgements.}
Zhengyi Yang is supported by Enmotech Data AU.
We would like thank the anonymous reviewers at ADC2023 Shepherding Track for their valuable insights and suggestions that significantly contributed to the paper's improvement.

\clearpage

\bibliographystyle{ieeetr}
\bibliography{ref}

\begin{thebibliography}{10}

\bibitem{ADC2023_JIT}
M.~Ma, Z.~Yang, K.~Hao, L.~Chen, C.~Wang, and Y.~Jin, ``An empirical analysis
  of just-in-time compilation in modern databases,'' in {\em Databases Theory
  and Applications} (Z.~Bao, R.~Borovica-Gajic, R.~Qiu, F.~Choudhury, and
  Z.~Yang, eds.), (Cham), pp.~227--240, Springer Nature Switzerland, 2024.

\bibitem{1960article}
J.~McCarthy, ``Recursive functions of symbolic expressions and their
  computation by machine,'' {\em Commun ACM}, vol.~3, 01 1960.

\bibitem{neumann2011efficiently}
T.~Neumann, ``Efficiently compiling efficient query plans for modern
  hardware,'' {\em Proceedings of the VLDB Endowment}, vol.~4, no.~9,
  pp.~539--550, 2011.

\bibitem{PostgreSQL}
{PostgreSQL.)}, ``Postgresql,'' 2023, June 8.
\newblock \url{https://www.postgresql.org/}.

\bibitem{Mutable}
{M. O.)}, ``Github - mutable-org/mutable: A database system for research and
  fast prototyping,'' 2023, June 7.
\newblock \url{https://github.com/mutable-org/mutable/}.

\bibitem{AOTvsJIT}
A.~Wade, P.~Kulkarni, and M.~Jantz, ``Aot vs. jit: impact of profile data on
  code quality,'' {\em ACM SIGPLAN Notices}, vol.~52, pp.~1--10, 06 2017.

\bibitem{IL}
{(Margaret Rouse, A. P.)}, ``Intermediate language. techopedia.,'' 2011,
  November 16.
\newblock
  \url{https://www.techopedia.com/definition/24290/intermediate-language-il-net}.

\bibitem{METULA20113}
E.~Metula, ``Chapter 1 - introduction,'' in {\em Managed Code Rootkits}
  (E.~Metula, ed.), pp.~3--21, Boston: Syngress, 2011.

\bibitem{tpch}
{( TPC-H Homepage. (n.d.).)}, ``Tpc-h homepage..''
\newblock \url{ https://www.tpc.org/tpch/}.

\bibitem{Volcano}
G.~Graefe, ``Volcano—an extensible and parallel query evaluation system,''
  {\em Knowledge and Data Engineering, IEEE Transactions on}, vol.~6, pp.~120
  -- 135, 03 1994.

\bibitem{Hyper-Pipelining}
P.~Boncz, M.~Zukowski, and N.~Nes, ``Monetdb/x100: Hyper-pipelining query
  execution,'' {\em 2nd Biennial Conference on Innovative Data Systems
  Research, CIDR 2005}, 01 2005.

\bibitem{holistic}
K.~Krikellas, S.~Viglas, and M.~Cintra, ``Generating code for holistic query
  evaluation,'' pp.~613--624, 01 2010.

\bibitem{Hyper}
{Muehlbauer, T.)}, ``Hyper: Hybrid oltp\&olap high-performance database system.
  hyper: Hybrid oltp\&olap high-performance database system.,'' n.d.
\newblock \url{https://hyper-db.de/}.

\bibitem{leis2014morsel}
V.~Leis, P.~Boncz, A.~Kemper, and T.~Neumann, ``Morsel-driven parallelism: a
  numa-aware query evaluation framework for the many-core age,'' in {\em
  Proceedings of the 2014 ACM SIGMOD international conference on Management of
  data}, pp.~743--754, 2014.

\bibitem{kohn2018adaptive}
A.~Kohn, V.~Leis, and T.~Neumann, ``Adaptive execution of compiled queries,''
  in {\em 2018 IEEE 34th International Conference on Data Engineering (ICDE)},
  pp.~197--208, IEEE, 2018.

\bibitem{haffner2023simplified}
I.~Haffner and J.~Dittrich, ``A simplified architecture for fast, adaptive
  compilation and execution of sql queries,'' {\em Proc EDTB}, 2023.

\bibitem{python}
{S.}, ``Python jit compilers,'' 2023, June 21.
\newblock [Online; accessed 2023, June 21].

\bibitem{LUAJIT}
{The LuaJIT Project. (n.d.)}, ``The luajit project.''

\bibitem{inproceedings}
T.~Cwi and P.~Boncz, ``Exploring query execution strategies for jit,
  vectorization and simd,'' 06 2017.

\bibitem{LLVMJIT}
postgresql-llvmjit Fedora Packages.~(n.d.)., ``Postgresql-llvmjit - fedora
  packages.,'' ().
\newblock
  \url{https://packages.fedoraproject.org/pkgs/postgresql/postgresql-llvmjit/}.

\bibitem{loopflatten}
{S. Pop, R. Yazdani, and Q.}, ``Improving gcc’s auto-vectorization with
  if-conversion and loop flattening for amd’s bulldozer processors,'' {\em in
  GCC Developers’ Summit}, p.~89, 01 2010.

\bibitem{graefe1994volcano}
G.~Graefe, ``Volcano/spl minus/an extensible and parallel query evaluation
  system,'' {\em IEEE Transactions on Knowledge and Data Engineering}, vol.~6,
  no.~1, pp.~120--135, 1994.

\bibitem{Binaryen}
{W. (n.d.)}, ``binaryen/readme.md at main,'' 2017.
\newblock \url{https://github.com/WebAssembly/binaryen}.

\bibitem{V8}
{Liftoff: a new baseline compiler for WebAssembly in V8 }, ``Liftoff: A new
  baseline compiler for webassembly in v8 · v8,'' 2018, August 20.
\newblock \url{ https://v8.dev/blog/liftoff}.

\bibitem{DeepinTPCH}
P.~Boncz, T.~Neumann, and O.~Erling, ``Tpc-h analyzed: Hidden messages and
  lessons learned from an influential benchmark,'' pp.~61--76, 01 2014.

\bibitem{boncz2014tpc}
P.~Boncz, T.~Neumann, and O.~Erling, ``Tpc-h analyzed: Hidden messages and
  lessons learned from an influential benchmark,'' in {\em Performance
  Characterization and Benchmarking: 5th TPC Technology Conference, TPCTC 2013,
  Trento, Italy, August 26, 2013, Revised Selected Papers 5}, pp.~61--76,
  Springer, 2014.

\bibitem{BranchPrediction}
E.~Quiñones, J.-M. Parcerisa, and A.~González, ``Improving branch prediction
  and predicated execution in out-of-order processors,'' pp.~75 -- 84, 03 2007.

\bibitem{LLDB}
{LLDB.}, ``Lldb,'' (2023, September 4).
\newblock \url{https://lldb.llvm.org/}.

\bibitem{GDB}
{GDB: The GNU Project Debugger. (n.d.)}, ``Gdb: The gnu project debugger.,''
  ().
\newblock \url{https://www.sourceware.org/gdb/}.

\bibitem{SIMD}
{W. (n.d.)}, ``relaxed-simd/proposals/simd/simd.md at main ·
  webassembly/relaxed-simd. github..''
\newblock
  \url{https://github.com/WebAssembly/relaxed-simd/blob/main/proposals/simd/SIMD.md}.

\bibitem{FusedTableScans}
M.~Dreseler, J.~Kossmann, J.~Frohnhofen, M.~Uflacker, and H.~Plattner, ``Fused
  table scans: Combining avx-512 and jit to double the performance of
  multi-predicate scans,'' pp.~102--109, 04 2018.

\bibitem{HyPerArchitecture}
A.~Kemper and T.~Neumann, ``Hyper: A hybrid oltp\&olap main memory database
  system based on virtual memory snapshots,'' pp.~195--206, 04 2011.

\bibitem{Umbra}
T.~Kersten, V.~Leis, and T.~Neumann, ``Tidy tuples and flying start: fast
  compilation and fast execution of relational queries in umbra,'' {\em The
  VLDB Journal}, vol.~30, 09 2021.

\bibitem{MonetDB}
M.~Zukowski, P.~Boncz, N.~Nes, and S.~Héman, ``Monetdb/x100 - a dbms in the
  cpu cache,'' {\em IEEE Data Eng. Bull.}, vol.~28, pp.~17--22, 01 2005.

\end{thebibliography}

\end{document}